\documentclass[12pt]{article}
\usepackage{amssymb}
\usepackage{epsfig}

\catcode`\@=11
\def\numberbysection{\@addtoreset{equation}{section}
        \def\theequation{\thesection.\arabic{equation}}}

\def\beq{\begin{equation}}
\def\eeq{\end{equation}}
\numberbysection
\begin{document}
\begin{titlepage}
\begin{center}
\hfill \\
\vskip 1.in {\Large \bf Bumblebee gravity with cosmological constant} \vskip 0.5in P. Valtancoli
\\[.2in]
{\em Dipartimento di Fisica, Polo Scientifico Universit\'a di Firenze \\
and INFN, Sezione di Firenze (Italy)\\
Via G. Sansone 1, 50019 Sesto Fiorentino, Italy}
\end{center}
\vskip .5in
\begin{abstract}
We show how to find exact black hole solutions in bumblebee gravity with cosmological constant including $BTZ$ black holes..
\end{abstract}
\medskip
\end{titlepage}
\pagenumbering{arabic}
\section{Introduction}

Einstein's gravity is based on Lorentz symmetry. Some models of gravity predict that the universe is not so symmetrical, but they admit a privileged direction. These new models, called bumblebee gravity, are effective field theories describing a vector field with vacuum expectation value that spontaneously breaks the Lorentz symmetry.

One of the most frequent applications of bumblebee gravity models is to give a possible explanation of dark energy - the phenomenon responsible for the accelerated expansion of the universe. To test this idea, in \cite{1} a Schwarzschild-like black hole solution has been obtained. The extension to the Kerr metric is not so easy, and so far it has been possible to derive only a solution in the slow rotation limit ( see \cite{2}-\cite{3}).
In \cite{4} the black hole solution has been extended to the case of cosmological constant in ($3+1$) dimensions and moreover it has been found that the shadow of the black hole can be measurably smaller than in Einstein gravity.

In this work we produce several exact solutions of black hole with cosmological constant including the black hole $BTZ$ in ($2+1$) dimensions, solving exactly all the equations of motion. We also introduce an alternative solution method by directly working on the action of bumblebee gravity. The article is organized as follows. Let us first recall how to derive the black hole solution with cosmological constant of Einstein gravity directly from the action. Then we apply the same method to derive the bumblebee-deformed black hole solutions in both ($3+1$)d and ($2+1$)d. In both cases we find that it is necessary to introduce a linear potential in the bumblebee's field.

\section{AdS gravity in four dimensions}

To test our solution method, let us recall how to derive the black hole solution in Einstein gravity with cosmological constant in ($3+1$)d directly from the action

\beq S \ = \ \int \ d^4 x \ \sqrt{-g} \ [ \ R - 2 \Lambda \ ] \label{21} \eeq

To simplify the derivation, an ansatz is introduced for the metric

\beq g_{\mu\nu} \ = \ - e^{2\gamma} \ dt^2 \ + \ e^{2\rho} \ dr^2 \ + \ r^2 d \theta^2 \ + \ r^2 sin^2 \theta d \phi^2 \label{22}
\eeq

If we introduce the notation

\beq N^2(r) \ = \ e^{  2 \gamma  +  2 \rho} \label{23} \eeq

then the determinant of the metric has the following simple form

\beq \sqrt{-g} \ = \ N(r) \ r^2 \ sin \theta \label{24} \eeq

The Ricci tensor for the metric (\ref{22}) has as components

\begin{eqnarray}
R_{tt} & = & e^{2(\gamma-\rho)} \ \left[ \gamma_{/rr} \ + \ \gamma_{/r} \left( \ \gamma_{/r} \ - \ \rho_r \ + \ \frac{2}{r} \ \right) \right] \nonumber \\
R_{rr} & = & - \ \gamma_{/rr} \ + \ \gamma_{/r} \rho_{/r} \ - \ ( \gamma_{/r} )^2 \ + \ \frac{2}{r} \ \rho_{/r} \nonumber \\
R_{\theta\theta} & = & e^{-2 \rho} \ [ \ r ( \rho_{/r} - \gamma_{/r} ) \ - \ 1 \ ] \ + \ 1 \nonumber \\
R_{\phi\phi} & = & sin^2 \theta \ R_{\theta\theta} \label{25}
\end{eqnarray}

We also need to calculate the curvature based on the metric (\ref{22})

\beq R \ = \ 2 e^{-2 \rho} \ \left[ - \gamma_{/rr} \ - \ ( \gamma_{/r} )^2 \ + \ \gamma_{/r} \rho_{/r} \ - \ \frac{2}{r} \ \gamma_{/r} \ + \ \frac{2}{r} \ \rho_{/r} \ - \ \frac{1}{r^2} \right]
\ + \ \frac{2}{r^2} \label{26} \eeq

Consequently the action can be written as

\begin{eqnarray} S & = & 2 \int \ d^4 x \ N(r) r^2 sin \theta \left\{ e^{-2\rho} \left[ - \gamma_{/rr} \ - \ ( \gamma_{/r} )^2 \ + \ \gamma_{/r} \rho_{/r} \ - \ \frac{2}{r} \ \gamma_{/r} \ + \ \frac{2}{r} \ \rho_{/r} \right]
\right. \nonumber \\
 & + & \left. \frac{1}{r^2} \ - \Lambda \right\} \label{27} \end{eqnarray}

We reduce the dependence of the action to only two independent variables $ N(r)$ and $ e^{2\gamma} $ by substituting

 \beq e^{-2\rho} \ = \ \frac{e^{2\gamma}}{N^2(r)} \label{28} \eeq

Integrating by parts gives the final expression:

 \beq S \ = \ 2 \int \ d^4 x sin \theta \left\{ ( \ r ( e^{2 \gamma} )_{/r} \ + \ e^{2 \gamma} \ ) \ \frac{1}{N(r)} \ + \ N(r) \ ( \ 1 - \Lambda r^2 \ )
 \right\} \label{29} \eeq

 Now deriving the equations of motion for $ N(r)$ and $ e^{2\gamma} $ is straightforward:

 \begin{eqnarray}
 \delta{N} & \rightarrow &  1 - \Lambda r^2 \ = \ \frac{1}{N^2(r)} \ ( \ r ( e^{2 \gamma} )_{/r} \ + \ e^{2 \gamma} \ ) \nonumber \\
 \delta( e^{2\gamma} ) & \rightarrow & r \frac{d}{dr} \ \left[ \frac{1}{N(r)} \right] \ = \ 0  \label{210} \end{eqnarray}

 whose solution is

 \begin{eqnarray}
 N(r) & = & 1 \nonumber \\
 e^{2\gamma} & = & 1 \ - \ \frac{2M}{r} \ - \frac{\Lambda}{3} r^2 \label{211} \end {eqnarray}

As we have seen, directly working on the action can greatly simplify the calculation.

\section{AdS gravity with the bumblebee field}

In bumblebee gravity we introduce a bumblebee vector field $ b_\mu $ with a non-zero vacuum expectation value $ b_0 $ which leads to the spontaneous breaking of the Lorentz symmetry. The action of bumblebee gravity in the presence of the cosmological constant admits a simple solution (as also discussed in \cite{4}) only if we add the constraint

\beq b^\mu b_\mu \ = \  (b_0)^2 \label{31} \eeq

with a linear potential as follows

\begin{eqnarray} S & = & \int \ d^4 x \ \sqrt{-g} \left\{ \ R \ - \ 2 \Lambda \ + \ \xi \ b^\mu b^\nu R_{\mu\nu} \ - \ \frac{1}{4} \ b^{\mu\nu} b_{\mu\nu} \ +
\right. \nonumber \\
 & + & \left. \lambda \ \xi \ [ \ b^\mu b_\mu \ - \  (b_0)^2 \ ] \
\right\} \label{32} \end{eqnarray}

As solution ansatz we choose as usual

\beq b_\mu \ = \ ( 0, b_r ( r ), 0 , 0 ) \label{33} \eeq

It follows that $ b_{\mu\nu} \ = \ 0 $. We then define the parameter $l$

\beq l \ = \ \xi (b_0)^2 \label{34} \eeq

By substituting the ansatz and integrating by parts, the following formula is obtained

\begin{eqnarray} S & = & 2 \int \ d^4 x \ sin \theta \left\{
\left( \ r ( e^{2\gamma} )_{/r} \ + \ e^{2\gamma} \ \right) \ \frac{1}{N(r)} \ \left( \ 1 \ + \xi \left( \frac{b_r e^\gamma}{N(r)}
\right)^2 \ \right)  + \right. \nonumber \\
& + & \frac{\xi}{2} \ \frac{r^2 e^{2\gamma}}{N(r)} \left( \ \gamma_{/r} \ + \frac{2}{r} \ \right) \frac{d}{dr} \left( \frac{b_r e^\gamma}{N(r)}
\right)^2 \ + \ N(r) \ ( \ 1 \ - \ \Lambda r^2 ) + \nonumber \\
& + & \left. \frac{\lambda \ \xi}{2} \ r^2 \  N(r) \ \left[ \ \left( \ \frac{b_r e^\gamma}{N(r)}
\right)^2 - (b_0)^2  \ \right] \right\}
\label{35} \end{eqnarray}

The corresponding equations of motion can be decomposed into the following blocks

\begin{eqnarray}
\delta( e^{2\gamma} ) & \rightarrow & r \frac{d}{dr} \ \left( \ \frac{1}{N(r)} \ \right) \ = \ 0 \nonumber \\
\delta( e^{2\gamma)} ), \delta{N}, \delta{b_r} & \rightarrow & \lambda \ = \ \frac{1}{2 \ N^2 (r)} \ \frac{1}{r^2} \ \frac{d}{dr} \left( \ r^2 ( e^{2\gamma} )_{/r} \ \right) \nonumber \\
\delta{N} & \rightarrow & 1 - \Lambda r^2 \ = \ \frac{1+l}{N^2(r)} \ \left( \ e^{2\gamma} \ + \ r ( \ e^{2\gamma} \ )_{/r} \ \right) \nonumber \\
\delta{\lambda} & \rightarrow & b_r \ = \ b_0 \ e^{-\gamma} N(r)
\label{36} \end{eqnarray}

which are solved by

\begin{eqnarray}
e^{2\gamma} & = & 1 \ - \ \frac{2M}{r} \ - \ \frac{\Lambda}{3} \ r^2 \nonumber \\
N^2 ( r) & = & l \ + \ 1 \nonumber \\
\lambda & = & - \frac{\Lambda}{l+1} \nonumber \\
b_r & = & b_0 e^{-\gamma} \ N(r)
\label{37} \end{eqnarray}

 Finally, for this solution the following identities hold

\begin{eqnarray}
R_{tt} & = & \frac{\Lambda}{l+1} \ g_{tt} \nonumber \\
R_{rr} & = & \frac{\Lambda}{l+1} \ g_{rr} \nonumber \\
R_{\theta\theta} \ - \ \frac{l}{l+1} & = & \frac{\Lambda}{l+1} \ g_{\theta\theta} \nonumber \\
R_{\phi\phi} & = & sin^2 \theta \ R_{\theta\theta}
\label{38} \end{eqnarray}

We therefore obtain that the parameter $\lambda$ of the linear potential is different from zero at the level of the equations of motion and proportional to the cosmological constant. Only in the Schwarzschild case the linear potential can be omitted. We have then verified that the solution (\ref{37}) satisfies the equations of motion written with the traditional method.

\section{$BTZ$ black hole with the bumblebee field}

Also in this case to obtain a simple solution of the equations of motion it is necessary to add linearly the constraint (\ref{31}) to the action of the bumblebee gravity for the black hole $BTZ$, something that has not been considered in \cite{5}:

 \begin{eqnarray} S & = & \int \ d^3 x \ \sqrt{-g} \left\{ \ R \ - \ 2 \Lambda \ + \ \xi \ b^\mu b^\nu R_{\mu\nu} \ - \ \frac{1}{4} \ b^{\mu\nu} b_{\mu\nu} \ +
\right. \nonumber \\
 & + & \left. \lambda \ \xi \ [ \ b^\mu b_\mu \ - \  (b_0)^2 \ ] \
\right\} \label{41} \end{eqnarray}

The ansatz for the metric and the bumblebee field are as follows

\begin{eqnarray}
ds^2 & = & - \ e^{2\gamma} dt^2 \ + \ e^{2\rho} dr^2 \ + \ r^2 \ [ \ \Omega (r) dt + d\phi \ ]^2 \nonumber \\
b_\mu & = & ( 0, b_r (r), 0 )
\label{42} \end{eqnarray}

In this case the determinant of the metric is given by

\beq \sqrt{-g} \ = \ N(r) \ r \label{43} \eeq

The angular part of the metric greatly complicates the calculation of the Ricci tensor:

\begin{eqnarray}
R_{tt} & = & e^{2( \gamma -\rho)} \left[ \ \gamma_{/rr} \ + \ \gamma_{/r} \ ( \ \gamma_{/r} \ - \ \rho_{/r} \ ) \ + \ \frac{\gamma_{/r}}{r} \
\right] \ + \ r \ e^{-2\rho} \ \Omega^2 ( \rho_{/r} \ - \ \gamma_{/r} \ ) \ - \nonumber \\
& - & 3 r \ e^{-2\rho} \ \Omega \ ( \Omega_{/r} ) \ + \ r^2 \ e^{-2 \rho} \ \Omega \ ( \Omega_{/r} ) \ ( \ \gamma_{/r} \ + \ \rho_{/r} \ ) \ - \ \frac{r^2}{2} \ e^{-2 \rho} \ ( \Omega_{/r} )^2 \ - \nonumber \\
& - & r^2 \ e^{-2 \rho} \ \Omega \ ( \Omega_{/rr} ) \ - \ \frac{r^4}{2} \ e^{-{2\gamma -2 \rho}} \ \Omega^2 \ ( \Omega_{/r} )^2
\label{44} \end{eqnarray}

\begin{eqnarray}
R_{t \phi} & = & - \ \frac{r^2}{2} \ e^{-2\rho} \ ( \Omega_{/rr} ) \ - \ \frac{3r}{2} \ e^{-2\rho} \ ( \Omega_{/r} ) \ + \ r \ e^{- 2 \rho } \ ( \rho_{/r} \ - \ \gamma_{/r} \ ) \ \Omega \ + \nonumber \\
& + & \frac{r^2}{2} \ e^{-2\rho} \ ( \ \gamma_{/r} \ + \rho_{/r} \ ) \ \Omega_{/r} \ - \ \frac{r^4}{2} \ e^{-2\gamma -2 \rho} \ \Omega \ ( \Omega_{/r} )^2
\label{45} \end{eqnarray}

\beq R_{\phi\phi} \ = \ r \ e^{-2 \rho} \ ( \ \rho_{/r} \ - \ \gamma_{/r} \ ) \ - \ \frac{r^4}{2} \ e^{-2\gamma -2 \rho} \ ( \Omega_{/r} )^2
\label{46} \eeq

\beq R_{rr} \ = \  - \ \gamma_{/rr} \ + \ \frac{1}{r} \ \rho_{/r} \ + \ \gamma_{/r} \ \rho_{/r} \ - ( \gamma_{/r} )^2 \ + \ \frac{r^2}{2} \ e^{- 2 \gamma} \ ( \Omega_{/r} )^2
\label{47} \eeq

Luckily the curvature has a simple dependency on $ \Omega $

\beq R \ = \ - 2 \ e^{-2 \rho} \ \left[ \ \gamma_{/rr} \ + \ ( \gamma_{/r} )^2 \ - \ \gamma_{/r} \ \rho_{/r} \  + \ \frac{\gamma_{/r} }{ r } \ - \ \frac{\rho_{/r} }{ r } \ \right] \ +
\ \frac{r^2}{2} \ e^{-2\gamma -2 \rho} \ ( \Omega_{/r} )^2
\label{48} \eeq

By substituting the ansatz and integrating by parts, the following formula is obtained

\begin{eqnarray}
S & = & \int \ d^3 x \left\{ \left( \ ( e^{2\gamma} )_{/r} \ + \ \frac{r^3}{2} \left( \Omega_{/r} \right)^2  \ \right) \ \frac{1}{N(r)} \ \left( \ 1 \ + \xi \left( \frac{b_r e^\gamma}{N(r)}
\right)^2 \ \right) + \right. \nonumber \\
& + & \frac{\xi}{2 N(r)} \ \left[ \ r ( e^{2\gamma} )_{/r} \ + \ 2 e^{2\gamma} \ \right] \ \frac{d}{dr} \left( \frac{b_r e^\gamma}{N(r)}
\right)^2 \ - \ 2 \Lambda r N(r) + \nonumber \\
 & + & \left. \lambda \ \xi \ r \  N(r) \ \left[ \ \left( \ \frac{b_r e^\gamma}{N(r)}
\right)^2 - (b_0)^2  \ \right] \right\}
\label{49} \end{eqnarray}

The corresponding equations of motion can be reduced to the following blocks

\begin{eqnarray}
\delta( e^{2\gamma} ) & \rightarrow & r \frac{d}{dr} \ \left( \ \frac{1}{N(r)} \ \right) \ = \ 0 \nonumber \\
\delta( e^{2\gamma)} ), \delta{N}, \delta{b_r} & \rightarrow & \lambda \ = \ \frac{1}{r \ N^2 (r)} \left\{  - \left[ \ (e^{2\gamma})_{/r} \ + \ \frac{r^3}{2} \ ( \Omega_{/r} )^2 \ \right] \ + \
\frac{1}{2} \ \frac{d}{dr} \ \left[ \ r \ ( \ e^{2\gamma} )_{/r} \ + \ 2 e^{2\gamma} \right] \right\}
 \nonumber \\
\delta{N} & \rightarrow &    \left( \ (e^{2\gamma})_{/r} \ + \ \frac{r^3}{2} \ ( \Omega_{/r} )^2 \ \right) \ \frac{l+1}{N^2(r)} \ = \ - \ 2 \ \Lambda \ r  \nonumber \\
\delta{\lambda} & \rightarrow & b_r \ = \ b_0 \ e^{-\gamma} N(r) \nonumber \\
\delta{\Omega} & \rightarrow & \frac{d}{dr} \left( \ r^3 \ \Omega_{/r} \ \right) \ = \ 0
\label{410} \end{eqnarray}

which can be solved in the same way as for the case ($3+1$)d:

 \begin{eqnarray}
e^{2\gamma} & = & - \Lambda \ r^2 - \ M \ + \ \frac{J^2}{4r^2} \nonumber \\
\Omega & = & - \frac{J}{2r^2} \nonumber \\
N^2(r) & = & l \ + \ 1 \nonumber \\
\lambda & = & - \frac{2\Lambda}{l+1} \nonumber \\
b_r & = & b_0 \ e^{-\gamma} N(r)
\label{411} \end{eqnarray}

Obviously in the case of the black hole $BTZ$ the cosmological constant $ \Lambda $ is negative.

For this solution the following identity holds

\beq R_{\mu\nu} \ = \ \frac{2\Lambda}{l+1} \ g_{\mu\nu} \label{412} \eeq

Also in this case we have a non-zero value of the parameter $ \lambda $ of the linear potential as in the previous case. Unlike the Kerr metric, the presence of angular momentum does not alter the integrability of the system.

\section{Conclusions}

In this work we have studied Einstein's gravity coupled to a bumblebee field in the presence of the cosmological constant. Both in ($3+1$)d and in ($2+1$)d we have found that it is necessary to introduce a linear potential in the bumblebee field to obtain the spontaneous breaking of the Lorentz symmetry. The parameter $\lambda$ that describes the linear potential gets a non zero value in the equations of motion proportional to the cosmological constant. We have introduced an alternative method to derive the equations of motion directly from the action of the bumblebee gravity. The ansatz for the metric and the bumblebee field turns out to give a well defined system of equations of motion, a property which is not trivial.

In ($2+1$)d we have found that the integrability of the system is not affected by the presence of angular momentum. The properties of the black hole $BTZ$ deformed with the bumblebee field have been discussed in \cite{5} and we do not repeat them here. For the Kerr metric in bumblebee gravity the choice of the right ansatz is not so trivial and remains to be investigated in more detail.

\end{document}